\documentstyle[prd,tighten,aps]{revtex}
\begin{document}
\draft

\twocolumn[\hsize\textwidth\columnwidth\hsize\csname
@twocolumnfalse\endcsname
\renewcommand{\theequation}{\thesection . \arabic{equation} }
\title{\bf Cosmological predictions from the Misner brane}

\author{Pedro F. Gonz\'{a}lez-D\'{\i}az}
\address{Centro de F\'{\i}sica ``Miguel Catal\'{a}n'',
Instituto de Matem\'{a}ticas y F\'{\i}sica Fundamental,\\ Consejo
Superior de Investigaciones Cient\'{\i}ficas, Serrano 121, 28006
Madrid (SPAIN)}
\date{August 24, 2000}

\maketitle

\begin{abstract}
Within the spirit of five-dimensional gravity in the
Randall-Sundrum scenario, in this paper we consider cosmological
and gravitational implications induced by forcing the spacetime
metric to satisfy a Misner-like symmetry. We first show that in
the resulting Misner-brane framework the Friedmann metric for a
radiation dominated flat universe and the Schwarzschild or
anti-de Sitter black hole metrics are exact solutions on the
branes, but the model cannot accommodate any inflationary
solution. The horizon and flatness problems can however be
solved in Misner-brane cosmology by causal and noncausal
communications through the extra dimension between distant
regions which are outside the horizon. Based on a semiclassical
approximation to the path-integral approach, we have calculated
the quantum state of the Misner-brane universe and the quantum
perturbations induced on its metric by brane propagation along
the fifth direction. We have then considered testable
predictions from our model. These include a scale-invariant
spectrum of density perturbations whose amplitude can be
naturally accommodated to the required value 10$^{-5}$ -
10$^{-6}$, and a power spectrum of CMB anisotropies whose
acoustic peaks are at the same sky angles as those predicted by
inflationary models, but having much smaller secondary-peak
intensities. These predictions seem to be compatible with COBE
and recent Boomerang and Maxima measurements.
\end{abstract}


\vskip2pc]

\renewcommand{\theequation}{\arabic{section}.\arabic{equation}}

\section{\bf Introduction}
\setcounter{equation}{0}

Extra dimensions have a long history in gravitational physics
since Kaluza [1] and Klein [2] first introduced a fifth
coordinate in Einstein general relativity to account for a
unified description of graviy and electromagnetism. Although
string and superstring theories have reached a great deal of
progress using and extending the Kaluza-Klein idea, the issue of
extra dimensions has never stirred theorists so much as some
work carried out by Horava, Witten, Randall and Sundrum has done
recently. Working in the realm of the ten-dimensional $E_8\times
E_8$ heterotic string theory, Horava and Witten first showed [3]
that this theory can be related to an eleven-dimensional theory
on the orbifold ${\bf R}^{10}\times{\bf S}^1/{\bf Z}_2$ in such
a way that whereas the standard model particles are confined to
the four-dimensional spacetime, gravitons propagate in the full
bulk space. By simplifying the Harova-Witten framework to five
dimensions, Randall and Sundrum then derived [4,5] two very
interesting models where there exist two branes placed on the
extra direction (which are identical to two domain walls with
oposite tensions) in five-dimensional anti-de Sitter spacetime.
In their first model [4], they assumed that we live in the
negative-tension brane and proposed a mechanism to solve the
hierarchy problem based on a small extra dimension. This
scenario may have the serious problem that gravity is repulsive
in the brane with negative tension [6]. Although this conclusion
might be the subject of debate, one at least can certainly say
that a negative tension on the physical brane brings some
interpretational difficulties to the model. The second
Randall-Sundrum model assumed [5] that we live in the
positive-tension brane, while the other brane is moved off to
infinity, so localizing gravity in one of the two three-branes
only. The problem with this scenario is that the field equations
in the positive-tension brane are nonlinear in the source terms
[7].

Thus, the second Randall-Sundrum model avoids any possible
physical effect from repulsive gravity, or at least negative
brane tension in the observable universe, but unfortunately
leads to noncoventional (i.e. non Friedmann) cosmology [7,8]
when matter in the positive-tension brane is isotropically and
homogeneously distributed, or nonconventional chargeless,
nonrotating black holes [9] when the matter in that brane is
assumed to collapse without rotating beyond its trapped surface.
Again some controversy can arise concerning the cosmological
difficulties of the second Randall-Sundrum model - in particular
it could be thought that the nonlinear source terms are all
irrelevant at any era at which we are able to do cosmology.
However, a pefectly standard Friedmann evolution appears to be a
most desirable property for any cosmological model. In spite of
the several attempts made in order to reconcile the
nonconventional cosmological bihaviour with standard Friedmann
scenario based on inserting a cosmological constant in the brane
universe [10-13], within the spirit of the Randall-Sundrum
approach [4,5], or on reinterpreting black-hole physics in the
brane world [9], it appears that the cosmological and collapse
scenarios resulting from any of the two Randall-Sundrum models
may have serious difficulties when comparing them with current
observations or theoretical requirements.

On the other hand, several authors have obtained inflationary
solutions for the three-brane worlds with non-trivial
configurations in the extra dimension [11,14-16]. Most of such
solutions are based on imposing different absolute values for
the tensions of the two branes, taking the resulting neat
tension as the source of the exponential expansion [15]. Of
course, the implicit main aim of all these inflationary brane
models is at solving standard cosmological puzzles such as
horizon and flatness problems. It is in this sense that
inflationary mechanisms operating in brane worlds may however be
regarded as superfluous. In fact, it has been shown by Chung and
Freese [17] that signals traveling along null geodesics on the
extra dimension in Randall-Sundrum spacetimes actually connect
distant points which otherwise are outside the horizon, so
solving as well the horizon and, eventually, the flatness
problems and whereby making unnecessary inflation as a
cosmological ingradient of the brane worlds. Recently, a new
approach which in a way can be considered as a combination of
the two randall-Sundrum models, has been also suggested [18] to
solve the above potential shortcomings of such models. Given a
five-dimensional spacetime with two domain walls with oposite
tensions placed on the fifth coordinate, in order to suitably
represent the universe we live in the new approach chooses
neither of the two branes individually, but both of them
simultaneously by imposing Misner symmetry and whereby allowing
nonchronal regions with closed timelike curves (CTC's) only in
the bulk.

The resulting cosmological scenario placed on the two-brane
system is then quite conventional: it just describes a standard
Friedmann universe in the radiation dominated era. After
exploring further the gravitational physics of this scenario, it
is also shown in the present paper that such a scenario can
quite naturally accommodate Schwarzschild or anti-de Sitter
black holes by simply allowing matter in the branes to collapse
without rotating, but not any kind of inflationary mechanism.
Once the conventional causally-generated gravitational behaviour
is recovered for matter in the branes, this paper aims at
exploring cosmological predictions from Misner-brane universe
by: (i) investigating the noninflationary connection mechanisms
between distant points (which otherwise are outside the horizon)
that can solve the horizon and flatness problems, (ii)
formulating the quantum state of the Misner-brane universe by
resorting to the semiclassical approximation of the Euclidean
path-integral formalism, (iii) deriving the spectra of initial
density perturbations and CMB anisotropies arising from quantum
fluctuations induced on the branes by propagating along the
fifth direction. We regard as the main results of the paper the
predictions of a scale-independent spectrum for primordial
density perturbations whose amplitude can easily accommodate the
requirement that the density contrast be smaller or
approximately equal to 10$^{-5}$ [19], and of a power spectrum
for CMB anisotropies which seems to fit recent data by Boomerang
[20] and Maxima [21] experiments better than any other proposed
models.

We outline the rest of the paper as follows. In Sec. II we
review and extend the results of Ref. [18], where the
Misner-brane model was introduced, and discuss some of its
physical implications, including spacetime propagation of
quantum fields, a calculation of the value of the Planck mass
and some solutions of the Klein-Gordon equation for small
gravitational fluctuations. The spherically symmetric collapse
of neutral matter in the branes which leads to conventional
Schwarzschild or anti-de Sitter black holes is studied in Sec.
III. Sec. IV deals with the quantum state of the Misner-brane
universe. We estimate it by resorting to the semiclassical
approximation to the Euclidean path-integral formalism. It is
obtained that the probability for this universe increases as the
parameters that determine the amplitude of the matter-density
fluctuations decrease. The quantum effects induced on the brane
spacetime by propagating them along the fifth coordinate are
estimated in Sec. V using as well a semiclassical approximation
to the path integral. The results from this calculation are then
employed to analyse the possible solution to the above-alluded
cosmological puzzles, and to obtain and discuss the spectra of
primordial density fluctuations and CMB anisotropies. Finally,
we summarize and conclude in Sec. VI.

\section{\bf Misner-brane cosmology}
\setcounter{equation}{0}

In this section we shall first review the spacetime structure of
the Misner-brane universe such as it was introduced in Ref.
[18], adding then some new material referred to the comparation
of such a spacetime with that of the original Randall-Sundrum
models and the physical consequences that may be drawn from it.
Let us start with a five-dimensional spacetime with the fifth
dimension, $\omega$, compactified on S$^{1}$, with
$-\omega_{c}\leq\omega\leq\omega_{c}$, and satisfying the
orbifold symmetry $\omega\leftrightarrow-\omega$. On the fifth
direction there are two domain walls, with the brane at
$\omega=0$ having positive tension and that at $\omega=\omega_c$
having negative tension. In order to represent the universe we
live in, we choose neither of the two branes on $\omega$
individually, but both of them simultaneously; that is to say,
we shall provide the fifth coordinate with a periodic character,
in such a way that the branes at $\omega=0$ and
$\omega=\omega_c$ are identified with each other, so that if one
enters the brane at $\omega=0$, one finds oneself emerging from
the brane at $\omega=\omega_c$, without having experienced any
tension. If we then set the brane at $\omega=0$ into motion
toward the brane at $\omega=\omega_c$ with a given speed $v$, in
units of the speed of light, our space would resemble
five-dimensional Misner space [22], the differences being in the
spatial topology and in the definition of time and the closed-up
extra direction which would also contract at a rate $v$. Then,
time dilation between the two branes would inexorably lead to
the creation of a nonchronal region which will start forming at
the future of a given chronology horizon [23].

\subsection{\bf Brane spacetime with Misner symmetry}

We shall first consider the metric of the five-dimensional
spacetime in terms of Gaussian coordinates centered e.g. on the
brane at $\omega=0$. If we assume the three spatial sections on
the branes to be flat, then such a metric can be written in the
form [24]
\begin{equation}
ds^2= c^2(\omega,t)\left(d\omega^2-dt^2\right)+
a^2(\omega,t)\sum_{j=2}^{4}dx_j^2 ,
\end{equation}
where if we impose the orbifold condition
$\omega\leftrightarrow-\omega$ [4,5,24], the scale factors $c$
and $a$ are given by
\begin{equation}
c^2(\omega,t)
=\frac{\dot{f}(u)\dot{g}(v)}{\left[f(u)+g(v)\right]^{\frac{2}{3}}}
,\;\;\; a^2(\omega,t) =\left[f(u)+g(v)\right]^{\frac{2}{3}} ,
\end{equation}
with $u=t-|\omega|$ and $v=t+|\omega|$ the retarded and advanced
coordinates satisfying the orbifold symmetry, where we have
absorbed some length constants into the definition of $t$ and
$\omega$, and the overhead dot denotes derivative with respect
to time $t$. If no further symmetries are introduced then $f(u)$
and $g(v)$ are arbitrary functions of $u$ and $v$, respectively
[24] . However, taking metric (2.1) to also satisfy the (Misner)
symmetry [25,26]
\[\left(t,\omega,x_2,x_3,x_4\right)\leftrightarrow\]
\[\left(t\cosh(n\omega_c)+\omega\sinh(n\omega_c),
t\sinh(n\omega_c)+\omega\cosh(n\omega_c),\right.\]
\begin{equation}
\left. x_2,x_3,x_4\right)
,
\end{equation}
where $n$ is any integer number, makes the functions $f(u)$ and
$g(v)$ no longer arbitrary. Invariance of metric (2.1) under
symmetry (2.3) can be achieved if we choose for $f(u)$ and
$g(v)$ the expressions $f(u)=Z_u\ln u+Y_u$, $g(v)=Z_v\ln v+Y_v$,
where the $Z$'s and $Y$'s are arbitrary constants. For the sake
of simplicity, throughout this paper we shall use the simplest
choice
\begin{equation}
f(u)=\ln u ,\;\;\; g(v)=\ln v .
\end{equation}
Imposing symmetry (2.3) together with the choice for the scale
factors given by expressions (2.4) fixes the topology of the
five-manifold to correspond to the identification of the domain
walls at $\omega=0$ and at $\omega=\omega_c$ with each other, so
that if one enters one of these branes then one finds oneself
emerging from the other.

The periodicity property on the extra direction can best be
explicited by introducing the coordinate transformation
\begin{equation}
\omega=T\sinh(W) ,\;\;\; t=T\cosh(W) ,
\end{equation}
with which metric (2.1) becomes
\begin{equation}
ds^2=
\frac{\left(\frac{\dot{T}^2}{T^2}-
\dot{W}^2\right)}{\ln^{\frac{2}{3}}T^2}\left(T^2
dW^2-dT^2\right)+ \ln^{\frac{2}{3}}T^2\sum_{j=2}^4 dx_j^2 .
\end{equation}
Although now metric (2.6) and the new coordinate
$T=\sqrt{t^2-\omega^2}$ (which is timelike provided that $\ln
T\geq const.\pm W$) are both invariant under symmetry (2.3), the
new extra coordinate $W$ transforms as
\begin{equation}
W\equiv\frac{1}{2}\ln\left(\frac{t+|\omega|}{t-|\omega|}\right)
\leftrightarrow W+n\omega_c
\end{equation}
under that symmetry. On the two identified branes making up the
Misner-brane universe, we can describe the four-dimensional
spacetime by a metric which can be obtained by slicing the
five-dimensional spacetime given by metric (2.6), along surfaces
of constant $W$, i.e.
\begin{equation}
ds^2= -\frac{\dot{T}^2}{T^2\ln^{\frac{2}{3}}T^2}dT^2+
\ln^{\frac{2}{3}}T^2\sum_{j=2}^4 dx_j^2 ,
\end{equation}
which will be taken throughout this paper to describe the
spacetime of the universe we live in. On the surfaces at
constant $T=T_c$, the resulting four-dimensional metric,
\[ds^2=-\frac{dW^2}{\sinh^{2}W\ln^{2/3}T_c^2}
+\ln^{2/3}T_c^2\sum_{j=2}^{4}dx_j^2 ,\] keeps a Lorentzian
signature even when we rotate to the consistent Euclidean sector
$W=i\Omega$ because on that sector time $t$ is still real (see
later on).

The energy-momentum tensor for the brane universe will now have
the form:
\begin{equation}
T_i^k= \frac{\delta(\omega-n\omega_c)}{c_b}{\rm
diag}\left(-\rho,p,p,p,0\right),\;\; n=0,1,2,3... ,
\end{equation}
where $c_b\equiv(t,\omega=n\omega_c)$. This tensor should be
derived using the Israel's jump conditions [27] that follow from
the Einstein equations. Using the conditions computed by
Bin\'{e}truy, Deffayet and Langlois [7] and the metric (2.6) we then
[24] obtain for the energy density and pressure of our
Misner-brane universe:
\begin{equation}
\rho= -\frac{4T\dot{W}}{\kappa_{(5)}^2\ln^{\frac{2}{3}}T^2
\left(|\dot{T}^2-T^2\dot{W}^2|\right)^{\frac{1}{2}}}
\end{equation}
\begin{equation}
p=\frac{2T\dot{T}^2\ln^{\frac{2}{3}}T^2}{\kappa_{(5)}^2
\left(|\dot{T}^2-T^2\dot{W}^2|\right)^{\frac{5}{2}}}
\frac{d}{dt}\left(\frac{T\dot{W}}{\dot{T}}\right)-\frac{1}{3}\rho .
\end{equation}
Thus, both the energy density $\rho$ and the pressure $p$,
defined by expressions (2.10) and (2.11), respectively,
identically vanish on the sections $W$=const. Therefore, taking
the jump of the component ($\omega,\omega$) of the Einstein
equations with the orbifold symmetry [7], one gets on the
identified branes
\begin{equation}
\frac{\dot{a}_b^2}{a_b^2} +\frac{\ddot{a}_b}{a_b}=
\frac{\dot{a}_b\dot{c}_b}{a_b c_b} ,
\end{equation}
where $a_b\equiv a(t,\omega=n\omega_c)$, with $n=0,1,2,3...$, is
the scale factor in our Misner-brane universe.

The breakdown of arbitrariness of functions $f(u)$ and $g(v)$
imposed by symmetry (2.3) prevents the quantity $c_b$ to be a
constant normalizable to unity, so the right-hand-side of Eq.
(2.12) can be expressed in terms of coordinates $T,W$ as:
\begin{equation}
\frac{\dot{a}_b\dot{c}_b}{a_b c_b}= -\frac{1+\frac{1}{3\ln
T}}{3T^2\cosh^2 W\ln T} .
\end{equation}
A simple dimensional analysis (performed after restoring the
constants absorbed in the definitions of $t$ and $\omega$ in
Eqs. (2.2)) on the right-hand-side of Eq. (2.13) indicates that
if this side is taken to play the role of the source term of the
corresponding Friedmann equation, then it must be either
quadratic in the energy density if we use
$\kappa_{(5)}^2=M_{(5)}^{-3}$ (with $M_{(5)}$ the
five-dimensional reduced Planck mass) as the gravitational
coupling, or linear in the energy density and pressure if we use
$\kappa_{(4)}^2=8\pi G_N=M_{(4)}^{-2}$ (with $M_{(4)}$ the usual
four-dimensional reduced Planck mass) as the gravitational
coupling. Since $\kappa_{(4)}^2$ should be the gravitational
coupling that enters the (Friedmann-) description of our
observable four-dimensional universe, we must choose the
quantity in the right-hand-side of Eq. (2.13) to represent the
combination $-\kappa_{(4)}^2(\rho_b+3p_b)/6$ which should be
associated with the geometrical left-hand-side part of Eq.
(2.12) of the corresponding Friedmann equation, when the term
proportional to the bulk energy-momentum tensor
$T_{\omega\omega}$ is dropped by taking the bulk to be empty. We
have then,
\begin{equation}
\rho_b+3p_b =\frac{2\left(1+\frac{1}{3\ln
T}\right)}{\kappa_{(4)}^2 T^2\cosh^2 W\ln T} .
\end{equation}
The four-dimensional metric (2.8) can be expressed as that of a
homogeneous and isotropic universe with flat spatial geometry,
$ds^2=-d\eta^2+a(\eta)_b^2\sum_{j=2}^4 dx_j^2$, if we take for
the cosmological time $\eta=3a(\eta)_b^2/(4\cosh
W)=3\ln^{2/3}T^2/(4\cosh W)$. In this case, the scale factor
$a(\eta)_b$ corresponds to that of a radiation dominated flat
universe, with $\cosh W=const$ expressing conservation of rest
energy, and $p_b=\rho_b/3$ at sufficiently small $\eta$. In
fact, for small $\eta$, it follows then from Eq. (2.14)
\[\rho_b\equiv \rho_b(T,\eta)\simeq\frac{4}{3\kappa_{(4)}^2
T^2\cosh^2(W)a(\eta)_b^6} ,\] or
\[\rho_b(\eta)=a(\eta)_b^2 T^2\cosh^2 W\rho_b(T,\eta)\simeq
\frac{3}{32\pi G_N\eta^2} ,\]
when expressed in terms of the cosmological time $\eta$ only.

At least in the noninflationary versions of the Randall-Sundrum
models, the tensions on the two branes are assumed to be given
by [4,5] $V_{\rm visible}=-V_{\rm hidden}=\rho_0=-p_0$, where
$\rho_0$ and $p_0$ are the energy density and pressure induced
from the bulk on the brane at $y=0$ ($\omega=0$). In the case of
the Misner-brane world, $\rho_0=p_0=0$, and hence $V_{\rm
visible}= V_{\rm hidden}=0$, so when one enters one brane one
finds oneself emerging from the other brane without having
experienced any tension. In this case, we have however a nonzero
"effective" tension which can be defined as $V\propto\rho_b >0$.
On the other hand, Cs\'{a}ki et {\it al.} have found [10] that any
problems arising from having a brane with negative tension
disappear in a radiation dominated universe, even in the
Randall-Sundrum models.

Having shown that the Misner-brane cosmology based on ansatz
(2.4) matches the standard cosmological evolution in the
radiation dominated era, we turn now to investigate the
nonchronal character of the spacetimes described by metric
(2.6). Nonchronal regions in such spacetimes can most easily be
uncovered if we re-define the coordinates entering this metric,
such that $Y=W-\ln T$ and $\Theta=T^2$. In terms of the new
coordinates, the line element (2.6) reads:
\begin{equation}
ds^2= -\frac{\left(\dot{Y}^2
+\frac{\dot{Y}\dot{\Theta}}{\Theta}\right)}{\ln^{\frac{2}{3}}\Theta}
\left(\Theta dY^2
+dYd\Theta\right)+\ln^{\frac{2}{3}}\Theta\sum_{j=2}^4 dx_j^2 .
\end{equation}
This metric is real only for $\Theta>0$ in which case $Y$ is
always timelike if $\dot{Y}>0$. One will therefore [26] have
closed timelike curves (CTC's) only in the bulk, provided
$\Theta>0$, $\dot{Y}>0$. There will never be CTC's in any of the
branes, that is the observable universe.

On the other hand, Singularities of metrics (2.6), (2.8) and
(2.15) will appear at $T=0$ and $T=1$. The first one corresponds
to $\omega=t=0$, and the second one to $\eta=0$, the initial
singularity at $Y=W$, $t^2=1+\omega^2$, in the radiation
dominated universe. We note that the source term
$-\kappa_{(4)}^2(\rho_b+3p_b)/6$ given by Eq. (2.14) also
diverges at these singularities. The geodesic incompleteness at
$T=1$ can be removed in the five-dimensional space, by extending
metric (2.6) with coordinates defined e.g by $X=\int
dW/\ln^{\frac{1}{3}}T^2-3\ln^{\frac{2}{3}}T^2/4$, $Z=\int
dW/\ln^{\frac{1}{3}}T^2+3\ln^{\frac{2}{3}}T^2/4$. Instead of
metric (2.6), we obtain then
\[ds^2=
\frac{2}{3}(Z-X)\times\]
\begin{equation}
\left\{\exp\left[\sqrt{\frac{8}{27}}
(Z-X)^{\frac{3}{2}}\right]\dot{X}\dot{Z}dXdZ +\sum_{j=2}^4
dx_j^2\right\} ,
\end{equation}
where one can check that whereas the singularity at $T=0$ still
remains, the metric is now regular at $T=1$. Since replacing $W$
for $Y$ in Eqs. (2.6) and (2.15) simultaneously leads to the
condition $Y=-\frac{1}{2}\ln T +const.$, and hence, by the
definition of $Y$, $Y=const$ and $W=const$ at $T=1$, one can
choose the singularities at $T=1$ (i.e. at the initial time at
which the brane system starts evolving along $T>1$) to nest
chronology horizons in the five-space. So, CTC's would only
appear in the bulk.

\subsection{\bf Physical implications}

If one considers a quantum field propagating in our spacetime,
then the renormalized stress-energy tensor $\langle
T_{\mu\nu}\rangle_{ren}$ would diverge at the chronology
horizons [28]. The existence of this semiclassical instability
would support a chronology protection also against the existence
of our universe model. However chronology protection can be
violated in situations which use an Euclidean continuation and
lead to a vanishing renormalized stress-energy tensor
everywhere, even on the chronology horizons. In order to convert
metric (2.16) into a positive definite metric, it is covenient
to use new coordinates $p, q$, defined by $X=p-q$, $Z=p+q$, or
$T^2=\exp\left[(4q/3)^{3/2}\right]$, $W^2=4p^2 q/3$. A positive
definite metric is then obtained by the continuation $p=i\xi$
which, in turn, implies $W=i\Omega$. Furthermore, using Eqs.
(2.5) we can also see that this rotation converts the extra
direction $\omega$ in pure imaginary and keeps $t$ and $T$ real,
while making the first two of these three quantities periodic
and leaving $T$ unchanged. Two {\it ans\"atze} can then be used
to fix the value of $P_{\Omega}$, the period of $\Omega$ in the
Euclidean sector. On the one hand, from the expression
$\exp(W)\rightarrow\exp(i\Omega)$ we obtain $P_{\Omega}=2\pi$, a
result that allows us to introduce a self-consistent Li-Gott
vacuum [29], and hence obtain $\langle
T_{\mu\nu}\rangle_{ren}=0$ everywhere. On the other hand, if we
take $\exp(p)\rightarrow\exp(i\xi)$, then we get $P_{\Omega}=
2\pi\ln^{1/3}T^2$. In this case, for an automorphic scalar field
$\phi(\gamma X,\alpha)$, where $\gamma$ represents symmetry
(2.3), $\alpha$ is the automorphic parameter [30], $0<\alpha
<1/2$, and $X=t,\omega,x^2,x^3,x^4$, following the analysis
carried out in [31,32], one can derive solutions of the field
equation $\Box\phi=\Box\bar{\phi}=0$ by demanding
$t$-independence for the mode-frequency. This amounts [32] to a
quantum condition on time $T$ which, in this case, reads $\ln
T^2=(n+\alpha)^3\ln T_0^2$, where $T_0$ is a small constant
time. The use of this condition in the Hadamard function leads
to a value for $\langle T_{\mu\nu}\rangle_{ren}$ which is again
vanishing everywhere [32]. This not only solves the problem of
the semiclassical instability, but can also regularize
expression (2.14) at $T=0$ and $T=1$:
\begin{equation}
\rho_b+3p_b=
\frac{2T_0^{-2(n+\alpha)^3}\left(1+\frac{1}{3(n+\alpha)^3\ln
T_0}\right)}{\kappa_{(4)}^2\cosh^2(W) (n+\alpha)^3\ln T_0} ,
\end{equation}
which can never diverge if we choose the constant $T_0$ such
that $\ln T_0\neq 0$.

In order to extract some physical consequences from the
Misner-symmetry based brane model discussed so far, one would
compare our five-dimensional metric (2.6) with the solution
obtained by Randall and Sundrum [4,5] which reads:
\begin{equation}
ds^2=e^{-2k|y|}\eta_{\mu\nu}dx^{\mu}dx^{\nu}+dy^2 ,\;\;
\mu,\nu=0,1,2,3 ,
\end{equation}
where $\eta_{\mu\nu}$ is Minkowski metric and $0\leq y\leq\pi
r_c$ is the fifth coordinate. Actually, metric (2.6) can be
written in the same form as for metric (2.18) if we re-define
coordinates $T$ and $W$ such that
\begin{equation}
\tau=\int\frac{\sqrt{\dot{T}^2- T^2\dot{W}^2}}{T\ln^{2/3}T^2}dT
,
\;\;r_c y=\int\frac{\sqrt{\dot{T}^2-
T^2\dot{W}^2}}{\ln^{2/3}T^2}dW ,
\end{equation}
where $\tau=\int d\eta/a(\eta)$ is the conformal time relative
to four-space. On each brane defined at $W=W_0$ we can readily
integrate the first of Eqs. (2.19) to yield
$\tau=3\ln^{1/3}T^2/(2\cosh W_0)$, that is we re-obtain for the
scale factor the consistent value $a(\tau)=\frac{2}{3}\cosh
W_0\tau$ that describes a radiation dominated universe in terms
of conformal time. We also notice that on surfaces of constant
$T=T_c$, the second of Eqs. (2.19) can also be integrated to
give
\[r_c y= \frac{i}{2\ln^{1/3}T_c^2}
\ln\left(\frac{\sqrt{T_c^2+\omega^2}-T_c}{\sqrt{T_c^2+\omega^2}+
T_c}\right) ,\] which in turn implies $\sinh W\sinh\left(r_c y
\ln^{1/3}T_c\right)=\pm 1$, and hence we have that on such
constant surfaces
\begin{equation}
|y|=\frac{1}{r_c\ln^{1/3}T_c^2}
\ln\left[\coth\left(\frac{W}{2}\right)\right] ,
\end{equation}
meaning that $W$ tends to infinity (zero) as $y$ goes to zero
(infinity). Actually, using the coordinate re-definitions
(2.19), metric (2.6) can be written as
\begin{equation}
ds^2=\ln^{2/3}T^2\eta_{\mu\nu}dx^{\mu}dx^{\nu}+r_c^2 dy^2 ,
\end{equation}
where $0\leq y\leq\pi$. This shows invariance under brane
permutation, $\omega=0\leftrightarrow\omega=\omega_c$, as the
two branes are defined for the same value of $T$, $T\geq 1$. One
can then introduce the gravitational action for our Misner-brane
model to be:
\[S_{\rm grav}=2M^3\int d^{4}x\int dy\sqrt{-g^{(5)}}R^{(5)} . \]
The four-dimensional graviton zero mode follows from metric
(2.21) by replacing the Minkowski metric $\eta_{\mu\nu}$ for a
four-dimensional metric $\bar{g}_{\mu\nu}(x)$. It is described
[5] by an effective action following from taking $y=r_c y$ and
substitution in the action
\begin{equation}
S_{\rm grav}=2M^{3}r_c\int d^{4}x\int dy\ln^{2/3}T^2\bar{R} .
\end{equation}
From this action we can derive the Planck mass [5]
\begin{equation}
M_p^2=2M^3\int_{0}^{\pi r_c}dy\ln^{2/3}T^2 ,
\end{equation}
or using $\exp(-2ky)=\ln^{2/3}T^2$ and replacing $\pi r_c$ for
$y$ in the upper integration limit,
\begin{equation}
M_p^2=\frac{M^3}{k}\left(1-\ln^{2/3}T^2\right) ,
\end{equation}
which reduces to $M_p^2=M^3/k$ at $T=1$, the initial moment (big
bang) of the evolution of the Misner-brane universe. Thus, the
initial time at $T=1$ corresponds to taking the brane at
$\omega=\omega_c$ away to infinite in the second Randall-Sundrum
model [5]. As the Misner-brane universe then slowly evolves
along $T>1$ values, $M_p^2$ will decrease according to Eq.
(2.24) while the distance between the two branes along $\omega$
decreases, as dictated by Misner symmetry. It follows that the
velocity of expansion of the observable universe must be
proportional to the speed at which the two branes approach to
one another when Misner symmetry is satisfied.

On the other hand, as expressed as a non-relativistic problem,
the Klein-Gordon equation for small gravitation fluctuations $h$
in the Misner-brane case can be written as [5]
\begin{equation}
\left[-\frac{1}{2}\partial_z^2+V(z)\right]\Phi(z)=m^2\Phi(z) ,
\end{equation}
where
\begin{equation}
V(z)=\frac{15z^2}{8(k|z|+1)^2} -\frac{3k}{2}\delta(z) ,
\end{equation}
with the new coordinate $z$ defined by
\[z=\frac{1}{k}\left(\left|\ln^{-1/6}T^2\right|-1\right) \]
\begin{equation}
\Phi(z)=\ln^{-1/6}T^2\Phi(y)
\end{equation}
\[h(x,z)=\ln^{-1/6}T^2 h(x,y)\]
In Misner-brane cosmology the use of Eqs. (2.27) allows us to
obtain for the wave function of the bound-state graviton zero
mode ($m=0$):
\begin{equation}
\Phi_0(T)=\frac{1}{k}\sqrt{\left|\ln T^2\right|} .
\end{equation}
We note that $\Phi_0(1)=0$, $\Phi(T>1)>0$ and $\Phi(0)=\infty$.
This means that the graviton is here strongly localized at the
farthest possible distance from the branes along the fifth
coordinate, and that it can only physically influence the
four-dimensional universe once this has started to evolve, but
not at the moment of its creation at $T=1$. This result might be
interpreted by considering that the structure of spacetime at
just the moment of big bang is not influenced by quantum
gravity.

An additional tower of continuous $m\neq 0$ $KK$ (Kaluza-Klein)
modes [5] also exist whose wave functions are given in terms of
Bessel functions $J$ and $Y$ [33]. For the Misner-brane scenario
these wave functions are
\begin{equation}
\Phi_{\it C}(T)=
\left(\frac{1}{k}\left|\ln^{-1/3}T^2\right|\right)^{1/2} {\it
C}_{2}\left(\frac{m}{k}\sqrt{\left|\ln^{-1/3}T^2\right|}\right)
,
\end{equation}
with ${\it C}=J,Y$. Far from the branes, as $T$ approaches 0,
$\Phi_{Y}\rightarrow\infty$ and $\Phi_{J}\rightarrow 0$, while
near the branes, as $T$ approaches 1, the wave functions reach
an oscillatory regime:
\begin{equation}
\Phi_{\it C}(T\rightarrow 1)
\simeq\sqrt{\frac{2}{\pi m}}
S\left(\frac{m}{k}\sqrt{\left|\ln^{-1/3}T^2\right|}-\frac{5}{4}\pi\right)
,
\end{equation}
with $S=\sin$ if ${\it C}=Y$ and $S=\cos$ if ${\it C}=J$. It is
interesting to notice that all the above wave functions satisfy
the boundary condition
\begin{equation}
\left.\partial_{z}\Phi_{\jmath}(z)\right|_{T=1}=0 ,
\end{equation}
where $\jmath=0, {\it C}$. One would regard Eq. (2.31) as
expressing the initial condition for the gravitational quantum
state of the Misner-brane universe, and note that it is
equivalent to the boundary condition on the regulator brane at
$y_c=\pi r_c$ when $r_c\rightarrow\infty$, in the second
Randall-Sundrum model [5].

We note finally that if the $T$-quantization $\ln
T^2=(n+\alpha)\ln T_0^2$ is adopted, then the argument of the
Bessel function in wave function (2.29) becomes
\[\sqrt{\left|\frac{m^2}{k^2(n+\alpha)\ln^{1/3}T_0^2}\right|}
.\] It would then follow that the graviton mass is quantized in
units of $\ln^{1/3}T_0^2$, so that
\begin{equation}
m^2=k^2(n+\alpha)\ln^{1/3}T_0^2 .
\end{equation}
In this case, one would replace the gravitonless singularity at
the big bang for an initial quantum state describing a graviton
characterized by a minimal nonzero mass
$m_0^2=k^2\alpha\ln^{1/3}T_0^2$.

Before closing this section, it is convenient to make clear some
points of the Miner-brane model where an incomplete
interpretation could lead to misundertanding. Thus, at first
sight, it could seem that Misner symmetry describes simple and
familiar spacetimes. Specifically one would believe this by
showing that Misner symmetry converts the five-dimensional
metric (2.1) into merely a reparametrization of the Kasner-type
solution [34]. However, the simple transformation
\[Q=\frac{1}{2}(f+g)=\ln T ,\;\;\; X=\frac{1}{2}(g-f)=W \]
converts metric (2.1) into
\[ds^2=(2Q)^{-2/3}\dot{u}\dot{v}\left(-dQ^2+dX^2\right)
+(2Q)^{2/3}\sum_{j=2}^{4}dx_j^2 ,\] which differs from a
Kasner-type metric by the factor
$\dot{u}\dot{v}=1-|\dot{\omega}|^2$ in the first term of the
right-and-side. This factor cannot generally be unity in the
five-dimensional manifold. On surfaces of constant $W=W_0$,
according to Eqs. (2.5), we have $\dot{\omega}=\tanh W_0$, so
$\dot{u}\dot{v}=\cosh^{-2}W_0$ which can only be unity for
$W_0=0$ that is on the brane at $\omega=0$. However, as it was
pointed out in the introductory paragraph of this section,
besides identifying the two branes according to Eq. (2.7), the
Misner approach also requires that the closed up direction
$\omega$ contracts at a given nonzero rate $d\omega_c/d\eta=
-v_0$ [23]. This in turn means that once the branes are set in
motion toward one another at the rate $v_0$, symmetry (2.7)
should imply that for constant $W_0$,
\[\frac{dW_0(0)}{d\eta}=0\leftrightarrow\frac{dW_0(\eta)}{d\eta}
-nv_0=0 ,\] so that $dW_0(\eta)/d\eta\neq 0$ if $n\neq 0$. In
this case, we have $\bigtriangleup
W_0(\eta)=n\int_0^{\eta}d\omega_c=n\bigtriangleup_{\eta}\omega_c$,
and hence $W_0(\eta)=W_0(0)+n\bigtriangleup_{\eta}\omega_c=
n\bigtriangleup_{\eta}\omega_c >0$, provided that we initially
set $W_0\equiv W_0(0)=0$. It follows that $\dot{u}\dot{v}$ can
only be unity on the brane at $\omega=0$ when $n=0$ (i.e. at the
very moment when the brane universe was created and started to
evolve. We note that if we substract the zero-point contribution
$\alpha\ln^{1/3}T_0^2$, the quantization of $T$ discussed above
amounts to the relation $\eta\propto n^2$ and, therefore,
initial moment at $\eta=0$ means $n=0$), taking on
smaller-than-unity values thereafter, to finally vanish as
$\eta,n\rightarrow\infty$. Thus, one cannot generally consider
metric (2.1) or metrics (2.6) and (2.8) to be reparametrizations
of the Kasner solution neither in five nor in four dimensions,
except at the very moment when brane at $\omega=0$ starts being
filled with radiation, but not later even on this brane.

We note that in the case that Kasner metric would exactly
describe our spacetime (as it actually happens at the classical
time origin, $T=1$, $n=0$), Misner identification reduces to
simply identifying the plane $W=X=0$ with $W=X=n\omega_c$, that
is identifying $W$ on a constant circle, which does not include
CTC's. This picture dramatically changes nevertheless once $n$
and $\eta$ become no longer zero, so that
$\dot{u}\dot{v}=\cosh^{-2}W_0
< 1$ and the metric cannot be expressed as a reparametrization
of the Kasner metric. In that case, there would appear a past
apparent singularity [actually, a past event (chronology)
horizon] at $T=1$ for observers at later times $\eta,n\neq 0$,
which is extendible to encompass nonchronal regions containing
CTC's, as showed before by using the extended metric (2.16).
Indeed, the particular value of $T$-coordinate $T=1$ measures a
quantum transition at which physical domain walls (three-branes)
with energy density $\rho_b$ created themselves, through a
process which can be simply represented by the conversion of the
inextendible physical singularity of Kasner metric [34] at
$T=1$, $n=0$ into the coordinate singularity of the Misner-brane
metric at $T=1$, relative to observers placed at later times
$\eta,n\neq 0$, which is continuable into a nonchronal region on
the bulk space.

On the other hand, since the energy density $\rho$ and pressure
$p$ on any of the two candidate branes vanish, one might also
think that, related to the previous point, we are actually
dealing with a world with no branes, but made up enterely of
empty space. The conversion of the field-equation term (2.13) in
a stress-energy tensor would then simply imply violation of
momentum-energy conservation. However, the existence of an event
(chronology) horizon which is classically placed at $T=1$ for
the five-dimensional spacetime amounts to a process of quantum
thermal radiation from vacuum, similar to those happening in
black holes or de Sitter space [35,36], which observers at later
times $\eta, n>0$ on the branes would detect to occur at a
temperature $\beta\propto\ln^{-1/3}T^2$, when we choose for the
period of $\Omega$ (which corresponds to the Euclidean
continuation of the {\it timelike} coordinate $W$ on
hypersurfaces of constant $T$)
$P_{\Omega}=2\pi\ln^{1/3}T^2\propto a$. Thus, for such
observers, the branes would be filled with radiation having an
energy density proportional to
$\ln^{-4/3}T^2\propto\eta^{-2}=\rho_b$ and temperature
$\propto\eta^{-1/2}$, i.e. just what one should expect for a
radiation dominated universe and we have in fact obtained from
Eq. (2.14). Observers on the branes at times corresponding to
$T>1$, $n\neq 0$ would thus interpret all the radiating energy
in the four-dimensional Misner-brane universe to come from
quantum-mechanical particle creation near an event horizon at
$T=1$.

Moreover, in order to keep the whole two-brane system
tensionless relative to a {\it hypothetical} observer who is
able to pass through it by tunneling along the fifth dimension
(so that when the observer enters the brane at $\omega=0$ she
finds herself emerging from the brane at $\omega=\omega_c$,
without having experienced any tension), one {\it must} take the
tension $V_{\omega=0}=\rho_b>0$ and the tension
$V_{\omega=\omega_c}=-\rho_b$, and therefore the total tension
experienced by the hypothetical observer,
$V=V_{\omega=0}+V_{\omega=\omega_c}$ will vanish. Given the form
of the energy density $\rho_b$, this necessarily implies that
current observers should live on just one of the branes (e.g. at
$\omega=0$) and cannot travel through the fifth direction to get
in the other brane (so current observers are subjected to
chronology protection [37]), and that, relative to the
hypothetical observer who is able to make that traveling, the
brane which she emerges from (e.g. at $\omega=\omega_c$) must
then be endowed with an antigravity regime with $G_N <0$ [6,10],
provided she first entered the brane with $G_N >0$ (e.g. at
$\omega=0$).

\section{\bf black holes on the branes}
\setcounter{equation}{0}

The branes can be defined as the hypersurfaces at
$\omega$=const., with the fifth direction given as
$d\omega=n_{\mu}dx^{\mu}$, where $n_{\mu}$ is the vector unit
normal to the four-manifold {\bf B} (the brane), with $x^{\mu}$
the coordinates [6]. This definition implies [34]
\begin{equation}
a^{\mu}=n^{\nu}\nabla_{\nu}n^{\mu} ,
\end{equation}
so that the five-dimensional metric (2.6) can be assumed to be
\begin{equation}
ds^2=q_{\mu\nu}dx^{\mu}dx^{\nu}+d\omega^2 ,
\end{equation}
where $q_{\mu\nu}=g_{\mu\nu}-n_{\mu}n_{\nu}$ is the induced
metric on the four-manifold {\bf B}. With this notation the
gravitational field equations for the identified two Misner
branes can be written as [6]:
\begin{equation}
^{(4)}G_{\mu\nu}= -\Lambda_4 q_{\mu\nu}+8\pi
G_N\tau_{\mu\nu}+\kappa_{(5)}^4\pi_{\mu\nu}-E_{\mu\nu} ,
\end{equation}
in which $\tau_{\mu\nu}$ is the energy-momentum tensor in the
branes, and
\begin{equation}
\Lambda_4=
\frac{1}{2}\kappa_{(5)}^2\left(\Lambda_5+
\frac{1}{6}\kappa_{(5)}^2\lambda^2\right)
\end{equation}
\begin{equation}
\pi_{\mu\nu}= -\frac{1}{4}\tau_{\mu\alpha}\tau_{\nu}^{\alpha}
+\frac{1}{12}\tau\tau_{\mu\nu}
+\frac{1}{8}q_{\mu\nu}\tau_{\alpha\beta}\tau^{\alpha\beta}
-\frac{1}{24}q_{\mu\nu}\tau^2
\end{equation}
\begin{equation}
G_N=\frac{\kappa_{(5)}^4\lambda}{48\pi} ,
\end{equation}
with $\lambda$ the brane tension.

In vacuum $\tau_{\mu\nu}=0$, and hence $\pi_{\mu\nu}=0$, so that
from Eq. (3.3)
\begin{equation}
^{(4)}G_{\mu\nu}=-\Lambda_4 q_{\mu\nu}-E_{\mu\nu} .
\end{equation}
If, following Dadhich, Maartens, Papadopoulos and Zezania [9],
we choose the bulk cosmological constant such that it satisfies
$\Lambda_{(5)}=-\kappa_{(5)}^2\lambda/6$ (or rather for
Misner-brane cosmology, setting $\Lambda_{(5)}=0$ and taking
into account that the total brane tension is zero), then we have
$\Lambda_{(4)}=0$, so that
\begin{equation}
^{(4)}G_{\mu\nu}=-E_{\mu\nu} ,
\end{equation}
where in general we have for a static vacuum
\begin{equation}
E_{\mu\nu}\propto
U\left(u_{\mu}u_{\nu}+\frac{1}{3}h_{\mu\nu}\right) +P_{\mu\nu} ,
\end{equation}
with $U$ an effective energy density on the brane which arises
from the free gravitational field in the bulk, $P_{\mu\nu}$ the
effective anisotropic stress coming also from the free
gravitational field in the bulk, and
$h_{\mu\nu}=q_{\mu\nu}+u_{\mu}u_{\nu}$, $u_{\mu}$ being the
chosen four-velocity field.

Since if no further symmetry other than $Z_2$-symmetry is
included $U$ and $P$ are generally nonzero, it was obtained in
[9] that black holes in the brane at $\omega$=0 has a
Reissner-Nordstrom like metric, with the "tidal charge" arising
from the gravitational effects of the fifth dimension playing
the role of an "electric charge" which does not exist
physically. This rather nonconventional behaviour of black holes
may be regarded as the counterpart in static spherically
symmetric vacuum of the nonlinear right-hand-side terms that
appear in the field equations that correspond to the homogeneous
and isotropic flat brane cosmology [7]. If we impose
nevertheless Misner symmetry {\it $\hat{M}$} to the induced
four-metric $q_{\mu\nu}$, then we have
\begin{equation}
{\it \hat{M}}q_{\mu\nu}=q_{\mu\nu} ,
\end{equation}
and in this case $q_{\mu\nu}$ would no longer be arbitrary. This
amounts to vanishing values for the quantities $U$ and $P$.
Thus, if $q_{\mu\nu}$ is made to satisfy Misner symmetry,
$E_{\mu\nu}=0$ and the field equations (3.7) strictly correspond
to the vacuum case:
\begin{equation}
^{(4)}G_{\mu\nu}=0 .
\end{equation}
If, in addition to Misner symmetry, we further impose spheric
symmetry and staticness to the induced metric $q_{\mu\nu}$, we
then obtain either the usual Schwarzschild solution
\begin{equation}
-q_{tt}=\frac{1}{q_{RR}}=1-\left(\frac{2M}{M_p^2}\right)\frac{1}{R}
,
\end{equation}
in the case that we choose $\Lambda_{(5)}=0$, or the
Schwarzschild-anti de Sitter (or -de Sitter) solution otherwise.

In order to re-express the four-dimensional Misner-brane metric
(2.8) as a Schwarzschild line element, one will introduce first
spheric symmetry in coordinates $x_j$'s (i.e. $x_j\rightarrow
r,\phi,\theta$), and then the change of variables
\begin{equation}
R=r\ln^{1/3}T^2=\frac{r}{2\gamma_R\left(1+\gamma_R\right)}
\end{equation}
\[t=\frac{\ln^{1/3}T^2}{2\cosh W_0} \left[\ln^{1/3}T^2
+\frac{3}{4}\gamma_R\left(2\ln^{1/3}T^2-1\right)\right]\]
\begin{equation}
=\frac{1+\frac{3}{2}\gamma_R
\left[1-\gamma_R\left(1+\gamma_R\right)\right]}{8\cosh
W_0\gamma_R^2\left(1+\gamma_R\right)^2} ,
\end{equation}
where
\begin{equation}
\gamma_R=\sqrt{1-\frac{2M}{R}}  ,
\end{equation}
with $M$ the black hole mass. We finally note that the new
coordinates $R$ and $t$ depend on $T$ and $r$ only. The
four-dimensional Misner-brane metric for vanishing cosmological
constant in the bulk becomes
\begin{equation}
ds^2=-\gamma_R^2 dt^2 +\frac{dR^2}{\gamma_R^2} +R^2 d\Omega_2^2
,
\end{equation}
in which $d\Omega_2^2$ is the metric on the unit two-sphere.
Thus, the past and future singularities correspond to $T=1$, the
event horizon to $T=\infty$, and the spatial infinity to
$T=\exp(1/96)$. In this way, we have succeeded in obtaining
conventional black holes in the Misner-brane framework. One
should interpret this as a proof of the consistency of that
framework.

\section{\bf The quantum state of the Misner-brane universe}
\setcounter{equation}{0}

In this section we will calculate the wave function of the
Misner-brane universe by using the semiclassical approach to the
Euclidean path-integral formalism [38]. The boundary initial
condition for this universe should be that it was not created
from nothing like in the no boundary or tunneling conditions
[39], but created itself [40] as corresponds to the spacetime
slicing at sections with $W=W_0$=const. of a five-dimensional
manifold whose fifth direction is filled with CTC's. Such an
initial condition is already incorporated in the
four-dimensional metric given by Eq. (2.8). This can be written
\begin{equation}
ds^2= -\frac{dT^2}{T^2\cosh^{2}W_0\ln^{2/3}T^2} +\ln^{2/3}T^2
ds_3^2 ,
\end{equation}
where $ds_3^2$ is the three-dimensional flat metric.

The path integral would be given as:
\[\psi\left[h_{ij}\right] =\int_C
d\mu\left[g_{ab}\right]\exp\left\{iS\left[g_{ab}\right]\right\}
,\] where $S\left[g_{ab}\right]$ is the Hilbert-Einstein
Lorentzian action for the Misner-brane universe with
four-dimensional metric $g_{ab}$, $a,b=0,1,2,3$ and $C$ is the
class of four-geometries which are bounded by a given
hypersurface on which they induce the set of data $h_{ij}$,
$i,j=1,2,3$, predicted by Misner-like symmetry (2.3). To obtain
a manageable expression for the wave function one should make
the rotation $W=i\Omega$, keeping $T$ real, and then introduce
the semicalssical approximation [39,41] to yield:
\[\psi\sim\exp\left[-S_E(A)\right]  ,\]
with $a$ the scale factor. The Hilbert-Einstein action
containing the necessary boundary term [38] is
\[S(T)= -\frac{1}{16\pi G_N}\int
d^{3}x\int_1^TdT'\sqrt{-g(T')}R(T')\]
\begin{equation}
+\frac{1}{8\pi G_N}\int d^{3}xTrK(T)\sqrt{h(T)} ,
\end{equation}
in which $TrK$ is the trace of the second fundamental form, and
$g$ and $h$ are the determinants of the four-metric and the
induced metric on the boundary, respectively. From metric (4.1)
we have for the scalar curvature, $R$, and the extrinsic
curvature, $K$,
\begin{equation}
R=-\frac{8}{3}\frac{\cosh^{2}W_0}{\ln^{4/3}T^2}
\end{equation}
\begin{equation}
TrK=2\frac{\cosh W_0}{\ln^{2/3}T^2} .
\end{equation}
Inserting Eqs. (4.3) and (4.4) into the action (4.2) and
integrating, we obtain
\begin{equation}
S(T)=\frac{M_p^2V_3}{2\pi}\cosh W_0\ln^{1/3}T^2 ,
\end{equation}
where $M_p=(8\pi G_N)^{-1/2}$ is the four-dimensional Planck
mass and $V_3=\int d^{3}x$ is the three-volume. In terms of the
cosmological time $\eta=3\ln^{2/3}T^2/(4\cosh W_0)$, action
(4.5) becomes:
\begin{equation}
S(\eta)=\frac{M_p^2 V_3}{\sqrt{3}\pi}\cosh^{3/2}W_0 \eta^{1/2} .
\end{equation}
The total action will be the sum of action (4.5) [or (4.6)] at
$\omega=0$ and at $\omega=\omega_c$. In the Euclidean formalism
where $T$ is kept real and $W=i\Omega$, if we set $W_0=0$ and
take the Li-Gott type period $P_{\Omega}=2\pi$, and hence a
unique $\Omega$-separation between the branes of $2\pi$, the
action on the two branes turns out to be
\begin{equation}
S_E(T)_{\omega=0}=-S_E(T)_{\omega=\omega_c} =\frac{M_p^2
V_3}{2\pi}\ln^{1/3}T^2 ,
\end{equation}
because the Newton constant changes sign on the brane at
$\omega=\omega_c$ [6,10] (see also Sec. II), and we have taken
into account the Misner identification of the two branes and the
orbifold symmetry. Hence, if we consider a Li-Gott type
self-consistent vacuum, the total Euclidean action $S_E^{\rm
total}=S_E(T)_{\omega=0}+S_E(T)_{\omega=\omega_c}=0$ and
therefore the semiclassical probability for the Misner-brane
universe will be
\begin{equation}
P=\psi^2\sim e^{-2S_E^{\rm total}}=1 .
\end{equation}
According to the discussion in Sec. II, however, in order to
avoid the singular character of both the renormalized
stress-energy tensor and the initial moment of the universe,
instead of a Li-Gott type Euclidean period $P_{\Omega}=2\pi$
[29], an Euclidean period
$P_{\Omega}=2\pi\ln^{1/3}T^2=2\pi(n+\alpha)\ln^{1/3}T_0^2$
should be used. In this case, however, there is an ambiguity in
the choice of $\Omega$-separation between branes. Taking into
account once again Misner and orbifold symmetries, so as the
feature that cosmological time $\eta\propto(n+\alpha)^2$ implies
an upper bound $\sqrt{\eta}$ for the possible values which the
integer $n$ may take on, in this case, one should in fact place
the two branes at any $\Omega$-separation given by
$\bigtriangleup\Omega=2\pi(n'+\alpha)\ln^{1/3}T_0^2$, where
$n'=0,1,2,...,n$. Assuming a normalization of the minimum time
such that $T_0^2=e$, we would have then
\begin{equation}
\bigtriangleup\Omega=2\pi(n'+\alpha)\leq P_{\Omega} ,\;\;\;
n'=0,1,2,...,n ,
\end{equation}
and hence we get for the total Euclidean action \[S_E^{\rm
total}(n)\]
\begin{equation}
=\frac{M_p^2 V_3}{\pi}(n+\alpha)\sin^{2}(\pi\alpha)
\equiv\frac{M_p^2 V_3}{\pi}a(\eta)\sin^2(\pi\alpha) ,
\end{equation}
which is defined in terms of $n$ through the scale factor
$a(\eta)$, but does not depend on $n'$, and hence on brane
separation. Thus, the ambiguity in the choice of brane
separation does not manifest in the final expression for the
wave function representing the quantum state of the Misner-brane
universe, a fact that physically suffice to ensure the necessary
invariance of the quantum state.

The above normalization of the minimum time $T_0$ is rather
conventional. It has been introduced merely for the sake of
simplicity in the equations and will be kept in what follows
unless otherwise stated. We note that this normalization implies
that if we assume any particular value or constraint on the
parameter $\alpha$, we are actually assuming that particular
value or constraint on the quantity $\alpha\ln^{1/3}T_0^2$. On
the other hand, the total action would only become zero whenever
the automorphic parameter $\alpha$ [30] is zero. However,
$\alpha$ is defined so that it can only take on nonzero values
which are smaller than 1/2. Moreover, if this parameter
vanished, then the stress-energy tensor would diverge on the
$\eta$-time ground state $n=0$ (see Sec. II). Moreover, if as
required by Misner symmetry we set the branes into a mutually
approaching motion at a Lorentzian speed $\beta$ [23] whose
absolute value is directly related to the velocity with which
galaxies are receding from one another (see Sec. II), then there
must exist a linear relation between the three-volume $V_3$ and
the Lorentzian volume on the extra dimension $V_1=\pi \ell_c$.
Taking this relation to be $V_1 M_p^{-2}=\pi V_3$, from the
known expression $M_p^2=M_{(5)}V_1$ [4,5] we finally achieve the
condition
\begin{equation}
\pi V_3 M_{(5)}^{3}=1 .
\end{equation}
Inserting condition (4.11) into the Euclidean action (4.10) we
obtain
\begin{equation}
S_E^{\rm total}=\frac{M_p
V_1}{\pi^2}(n+\alpha)\sin^{2}(\pi\alpha) .
\end{equation}
Thus, for $\alpha<<1$, we can approximate $S_E^{\rm total}\simeq
M_p V_1\alpha^3$ on the ground state $n=0$. In this case, the
probability for the universe would be very large not just in the
ground state, but also on the low-lying excited states. The
ansatz $V_1 M_p^{-2}=\pi V_3$ can in fact be related with the
Hubble law as follows. Since action (4.12) does not depend on
spacelike directions, the continuity equation for the
Misner-brane universe should read for $\alpha<<1$
\begin{equation}
\frac{dP}{d\eta}=0,\;\;\; P=\exp\left(-2M_p V_1
a(\eta)\alpha^2\right) ,
\end{equation}
in which $a(\eta)$ is the scale factor. From expressions (4.13)
it inmediately follows
\begin{equation}
\beta\equiv v_0=-\frac{dV_1}{d\eta}=V_1 H  ,
\end{equation}
where $v_0$ is the speed at which the Misner-like circumference
$2\pi r_c$ (with $r_c$ being the radion [42]) is contracting
(see Sec. II), and $H=\dot{a}/a$ is the Hubble constant. Eq.
(4.14) is in fact the Hubble law for the extra direction.

\section{\bf The cosmological quantum metric}
\setcounter{equation}{0}

We turn now to study the possible effects that the extra
dimension may cause on the metric of the four-dimensional
Misner-brane universe. In principle, these effects can be of two
different types. On the one hand, propagation along the fifth
direction will causally modify the spacetime on the branes and,
on the other hand, communication between the two branes or
different regions on each of these branes can be carried out
through processes involving CTC's that shortcut spacetime along
the fifth direction. All of these effects can be
quantum-mechanically described by using the Euclidean
path-integral formalism [38]. Actually, it would be natural to
calculate the quantum-mechanically disturbed four-dimensional
metric as that effective metric which results from including all
the above-mentioned effects from the fifth dimension in the
four-dimensional Misner-brane metric (2.8) by propagating the
quantum state of the undisturbed metric through the fifth
direction from the position of one brane to the position of the
other. This propagator can typically be given as a path integral
of the form:
\begin{equation}
\langle g_{ab},\omega=0|g_{ab},\omega=\omega_c\rangle =
\int d\mu[g_{\mu\nu}]\exp\left\{iS[g_{\mu\nu}]\right\} ,
\end{equation}
where $a$,$b$=0,1,...,3, $\mu,\nu$=0,1,...,4, $d\mu$ is the
integration measure, and $S[g_{\mu\nu}]$ is the five-dimensional
Lorentzian Hilbert-Einstein action which is given by
\[S[g_{\mu\nu}]=-\frac{1}{16\pi G_{(5)}} \int
d^{3}x\int_0^{W_c}dW\int dT\sqrt{-g^{(5)}}R^{(5)}\]
\begin{equation}
+\frac{1}{8\pi G_{(5)}}\int d^{3}x\int_0^{W_c}dW
TrK_{(4)}\sqrt{h^{(4)}} ,
\end{equation}
with $W_c=\sinh^{-1}(\omega_c/T)$ the value of the extra
$W$-coordinate at $\omega_c$, $G_{(5)}$ the five-dimensional
gravitational constant, $g^{(5)}$ and $h^{(4)}$ the determinants
of the five-dimensional metric and the spacelike metric induced
on the boundary whose second fundamental form is $K_{(4)}$,
respectively, and $R^{(5)}$ the five-dimensional scalar
curvature.

Such as it is expressed by Eqs. (5.1) and (5.2) this path
integral does not have any clear physical significance because
the values $\omega=0$ and $\omega=\omega_c$ ( and hence $W=0$
and $W=W_c$) merely express coordinate labels. This should be
related to the well-known fact that classical general relativity
does not allow the existence of well-defined degrees of freedom
[34]. However, the kind of "quantization" of the fifth dimension
[induced by Misner identification of branes at these values of
$\omega$ together with its "quasiharmonic" discretization (see
Sec. II)] which comes in addition to the quantization implied by
the usual notion of path integral, allows us to take such
$\omega$ values as the true physical quantities that determine
the dynamics and actually the very existence of the universe
itself. Taking into account this additional quantization, so as
Misner identification of the two branes and orbifold symmetry,
the physically meaningless path integral (5.1) can be in fact
converted into the physical quantity:
\[\langle g_{ab},W=0|g_{ab},\bigtriangleup W\rangle
=\]
\begin{equation}
\left.\int d\mu[g_{\mu\nu}]\right|_{W=W_0}
\exp\left\{\left.iS[g_{\mu\nu}]\right|_{W=W_0}\right\} ,
\end{equation}
with $\bigtriangleup W=i\bigtriangleup\Omega$ the brane
separation in the Lorentzian sector, and
\[\left.S[g_{\mu\nu}]\right|_{W=W_0}=\]
\[-\frac{\bigtriangleup W}{16\pi G_{(5)}} \int
d^{3}x\int
dT\sqrt{\left.-g^{(5)}\right|_{W=W_0}}\left(\left.R^{(5)}\right|_{W=W_0}\right)\]
\begin{equation}
+\frac{\bigtriangleup W}{8\pi G_{(5)}}\int d^{3}x
\left(\left.TrK_{(4)}\right|_{W=W_0}\right)\sqrt{\left.h^{(4)}\right|_{W=W_0}} ,
\end{equation}
where the quantities $\left.X\right|_{W=W_0}$ are the same as
$X$ but evaluated at $W=W_0=$const. once they have been computed
using $W$ as a variable.

To obtain a manageable, physical expression of this propagator
we now make [38] an Euclidean rotation with $W=i\Omega$ and the
physically most interesting brane separation $\bigtriangleup
W=i\bigtriangleup\Omega =2\pi i(n'+\alpha)$, so that the
integral (5.3) becomes an integral over
$\exp\left\{-\left.S_{E}[g_{\mu\nu}]\right|_{\Omega=\Omega_0}\right\}$,
with $\Omega_0=$const. In the semiclassical approximation where
\begin{equation}
\left|\psi_{ab},\Omega\right.\rangle\rightarrow
\psi\left(g_{ab},\Omega\right)
\end{equation}
\begin{equation}
\tilde{g}_{ab}\equiv\left.g_{ab}\right|_{\Omega=\Omega_0}\rightarrow
\frac{\delta}{\delta
\pi^{ab}}\left\{\ln\left[\psi\left(a_{ab},\Omega\right)\right]\right\}
,
\end{equation}
with $\psi\left(g_{ab},\Omega\right)$ a quasiclassical wave
function and $\pi^{ab}=\ln^{-4/3}T^2$ the momentum conjugate to
$g_{ab}$, we obtain for the wave function
\begin{equation}
\psi\left[g_{ab},2\pi(n'+\alpha)\right]\sim
\psi\left[g_{ab},0\right]
\exp\left\{-\left.S_{E}\left[g_{\mu\nu}\right]\right|_{\Omega=\Omega_0}\right\}
,
\end{equation}
from which one can obtain an approximate expression for the
four-dimensional effective metric in terms of the cosmological
time containing the effects from the fifth dimension by using
the operation (5.6). It is worth noticing that the metric
eigenstate (5.7) does depend on brane separation, a fact which
one should expect as it expresses the already mentioned relation
between universal expansion and variation of brane separation.

The Euclidean action can now be evaluated from the
five-dimensional metric components. These first produce
\begin{equation}
\left.R^{(5)}\right|_{W=W_0}=\frac{8}{3}\frac{\cosh^{2}W_0}{\ln^{\frac{4}{3}}T^2}
\end{equation}
\begin{equation}
\left.TrK^{(4)}\right|_{W=W_0}=-\frac{4}{3}\frac{\cosh^{2}W_0}{\ln^{\frac{2}{3}}T^2}
,
\end{equation}
and then inserting expressions (5.8) and (5.9) into the action
(5.4), performing the integration over $T$, rotating
$W=i\Omega$, and finally setting $W_0=i\Omega_0=0$ (without loss
of generality) and $\bigtriangleup\Omega=2\pi(n'+\alpha)$ as in
Sec. IV, it is obtained for the Euclidean action:
\begin{equation}
S_E=-\frac{V_{3}(n'+\alpha)}{8G_{(5)}}\ln\left(\ln T^2\right)
,\;\; n'=0,1,2,...,n .
\end{equation}
The disturbed quasiclassical wave function becomes then
\begin{equation}
\psi\left[g_{ab},2\pi(n'+\alpha)\right]
\sim\psi\left[g_{ab},0\right]
\exp\left\{\frac{V_{3}(n'+\alpha)}{8G_{(5)}}\ln\left(\ln
T^2\right)\right\} ,
\end{equation}
which depends on both $n'$ and $n$, the latter dependence taking
place through $\ln T$. Using the definition $8\pi
G_{(5)}=M_{(5)}$ and the condition (4.10), we can obtain the
eigenvalues of the disturbed effective metric $\tilde{g}_{ab}$
by applying operation (5.6) to the wave function (5.11). We get
\begin{equation}
\tilde{g}_{ab}=g_{ab}+(n'+\alpha)\ln^{\frac{4}{3}}T^2 ,
\end{equation}
where $g_{ab}=\ln^{2/3}T^2$ is the classical metric. Taking into
account that $\sqrt{\eta}=(n+\alpha)$, one can finally express
Eq. (5.12) in terms of the cosmological time $\eta$ as:
\begin{equation}
\tilde{g}_{ab}=g_{ab}+(n'+\alpha)\eta^2
=g_{ab}+(n'+\alpha)(n+\alpha)^4 ,
\end{equation}
with the integer number $n'=0,1,2,..,n$ the same as that was
introduced in Eq. (4.9), and we have disregarded here and
hereafter dimensionless factors of order unity in the
expressions for $a$, $g_{ab}$ and $h_{ab}$. Thus, depending on
the values of $\eta(n)$, $\alpha$ and $n'$, two regimes can be
distinguished for expression (5.13): (i) $\eta >> 1$, $n'>0$
where $\tilde{g}_{ab}\simeq(n'+\alpha)\eta^2$ which reaches a
maximum value $\eta^{5/2}$ at $n'=n$, and (ii) $\eta
<< 1$ (i.e. $n$=0) where the second term in the right-hand-side
of Eq. ((5.13) can be regarded as a perturbation, $h_{ab}$, to
metric $g_{ab}$; in the extreme case that $n'=n=0$, we get a
minimum value for this perturbation $h_{ab}=\eta^{5/2}$. In the
following subsections we shall discuss some cosmological
consequences from the Misner-brane model in these two extreme
cases.

\subsection{\bf Solution to the horizon and flatness puzzles}

Misner-brane cosmology cannot accommodate any realistic model of
inflation and therefore contends with it as an explanation for
the physics in the early universe. You can readily convince
yourself of this incompatibility if you consider that for
inflation to take place in a brane world it is necessary that
the absolute value of the tension on the two involved branes be
different [15], a requirement which is both conceptually and
operationally incompatible with Misner symmetry, as is the
necessity for brane inflationary models [15] that functions
$f(u)$ and $g(v)$ be both exponential.

Nevertheless, the existence of one extra direction makes
inflation unnecessary to solve the horizon and flatness puzzles.
Even for a fully causal behaviour along the extra dimension,
Chung and Frees have shown [17] that signals traveling along
null geodesics on extra space may connect distant points which
otherwise are outside the four-dimensional horizon. If spacetime
tunneling is moreover allowed to occur, then additional two-way
transmissions of signals between spacelike separated regions
would occur that permitted such regions to come into thermal
contact [43]. Misner-brane cosmology combines these two
mechanisms and, therefore, by itself provides a solution to the
standard cosmological puzzles.

A more technical argument comes from the form of metric (5.13)
in the extreme case of regime (i) where $n'=n>>1$, i.e.
$\eta>>1$. Here, the scale factor for the radiation-dominated
universe is given by $\tilde{a}(\eta)=\eta^{5/4}$. Thus, after
the Planck era, we have $\dot{\tilde{a}}(\eta)\equiv
d\tilde{a}(\eta)/d\eta=\eta^{1/4}$, and hence
\begin{equation}
\dot{\tilde{a}}(\eta_{\rm exit})>\dot{\tilde{a}}(\eta_1) ,\;\;\;
\eta_{\rm exit}>\eta_1 ,
\end{equation}
where $\eta_{\rm exit}$ is the time at the radiation-matter
transition, and $\eta_1$ is an early time which still occurs
much later than Planck era. The inequality (5.14) clearly solves
the cosmological horizon puzzle [44]. Since one can always write
this inequality as the equality
\[\dot{\tilde{a}}(\eta_{\rm exit})=\beta\dot{\tilde{a}}(\eta_1) ,\;\;\;
\beta>1  ,\]
the ratios of the energy density to the critical densities
$\Omega_{\rm exit}$ and $\Omega_1$ can be related by
\begin{equation}
\beta^{2}\left|\Omega_{\rm exit}-1\right|=
\left|\Omega_1-1\right|,
\end{equation}
so implying that $|\Omega-1|$ need not be initially set to a
very small value in order to insure small later values. This
also provides with a solution to the flatness puzzle [44]
because $\beta$ can take on very large values even during the
radiation-dominated era, so that $\left|\Omega_{\rm
exit}-1\right|$, and hence the current value
$\left|\Omega_0-1\right|$, get on very small values without
resorting to any fine tuning.

\subsection{\bf Origin of density perturbations}

Perhaps the greatest success achieved by inflationary models so
far be their prediction of a scale-invariant spectrum for
density perturbations [45]. Such a success cannot however hide
the rather dramatic feature that all realistic models of
inflation considered so far fail to predict a reasonably small
amplitude for the perturbation spectrum [46]. In the absence of
any inflationary mechanism, Misner-brane cosmology should by
itself produce primordial density perturbations which, in this
case, must be generated by quantum fluctuations of the spacetime
metric itself, rather than any matter field. This inmediately
connects with the extreme regime (ii) with $\eta << 1$, $n'=n=0$
for the disturbed metric (5.13). In this regime, $g_{ab}=\eta$
and $h_{ab}=\eta^{5/2}$. The fluctuation spectrum will be given
as a Fourier transform of $h_{ab}$ given by
\begin{equation}
h_k=3\sqrt{\frac{2}{\pi}} \int_0^{\pi/k_*
^{ab}}h^{ab}\exp\left(ig_{ab}k^{ab}\right)dg_{ab} ,
\end{equation}
where the momenta $k_* ^{ab}$ and $k^{ab}$ are mutually related
by $k^{ab}=a(\eta)k_* ^{ab}$ for $k$-modes that enter the Hubble
radius, i.e. $k^{ab}=2\pi Ha(\eta)$ at $\eta=\eta_{\rm exit}$
[46]. Inserting $h^{ab}=\sqrt{\eta}$ and $g_{ab}=\eta$ in Eq.
(5.16) and performing the integral after noticing that the
horizon size is $H^{-1}=2\eta$, we obtain for $n'=n=0$ and
$\alpha << 1$
\begin{equation}
k^{3/2}h_k\simeq 3\sqrt{\frac{2\alpha}{\pi}} .
\end{equation}
For this spectrum, the contribution of each interval from $k$ to
$\gamma k$ ($\gamma >1$) to the total dispersion,
\begin{equation}
D=4\pi\int_{k}^{\gamma k}dk k^2\left|h_k^2\right|\simeq
72\alpha\ln\gamma ,
\end{equation}
is scale-independent, as required by the Zel´dovich-Harrison
theory [47] and is also predicted by realistic inflationary
models [45].

We turn now to consider density perturbations and their
spectrum. For metric perturbations of the type given by Eq.
(5.13), the density contrast of the perturbations of matter
density (or temperature) is given by [48]
\begin{equation}
\left(\frac{\delta\rho}{\rho}\right)^{ab}=
\frac{a\dot{a}\dot{h}^{ab}-h^{ab}}{3\left(1+\dot{a}^2\right)}
\simeq\frac{1}{3}\sqrt{\eta} ,
\end{equation}
where $a=\sqrt{\eta}$. The amplitude of the spectrum of density
fluctuations will be then
\begin{equation}
\delta_k=3\sqrt{\frac{2}{\pi}}
\int_{0}^{\pi/k_*^{ab}}\left(\frac{\delta\rho}{\rho}\right)^{ab}
\exp\left(ig_{ab}k^{ab}\right)dg_{ab}.
\end{equation}
Inserting expression (5.19) in Eq. (5.20) and performing the
integral transform we finally obtain for $n'=n=0$
\begin{equation}
k^{3/2}\left|\delta_k\right|\sim\sqrt{\frac{2\alpha}{\pi}}
\ln^{1/6}T_0^2 ,
\end{equation}
where we have restored, for the moment, a generic value for
$T_0$. This is one of the main results of this work. It states
that in Misner-brane cosmology the density of perturbations has
the scale-invariant spectrum with enough a small amplitude. From
the bounds on the anisotropy of CMB we know [49] that
$\sqrt{\alpha}\ln^{1/6}T_0^2$ should then be smaller than
10$^{-4}$, with possibly a value between 10$^{-5}$ and
10$^{-6}$. This is quite a reasonable bound on
$\alpha\ln^{1/3}T_0^2$ that, in the present model, is also
compatible with the perturbative character of $h_{ab}$ (needing
$\ln^{5/3}T_0^2\alpha^5<<g_{ab}<<1$ for $n'=n=0$), and
corresponds to a high probability for the universe (see Sec.
IV). On the other hand, it was seen in previous sections that
the lowest possible bound for both $\alpha$ and $\ln T_0^2$ can
never be zero. We note furthermore that perhaps the presence of
baryons in the perturbations would place the minimum size in
such a way that the minimum and therefore most probable value of
the quantity $\alpha^{1/2}\ln^{1/6}T_0^2$ be close to the
required value between 10$^{-5}$ and 10$^{-6}$. It is in this
sense that the results obtained so far for Misner-brane
cosmology conform better to experiment than those obtained from
realistic models of inflation which all imply too large an
amplitude for density fluctuations [45,46]. In the next
subsection, it will be seen that the higher-dimensional
cosmological approach considered in this paper seems to produce
better predictions than inflation also for the spectrum of CMB
anisotropies.

\subsection{\bf CMB anisotropies}

The fluid-dynamical theory underlying primordial fluctuations of
background temperature [50] predicts the existence of
small-angle CMB anisotropies coming from the primeval ripples,
first observed by COBE [19], at the recombination time. These
anisotropies correspond to sound waves and show a power spectrum
that carries precise fundamental information about the origin of
fluctuations and the fate of the universe. There are two current
paradigms able to predict the power spectrum of CMB
anisotropies, inflation [51] and topological defects,
particularly cosmic strings and textures [52]. Whereas realistic
models of inflationary cosmology predict a power spectrum which
is "superluminarly" generated by perturbations that exceeded the
horizon size after crossing it at a common time, and contains a
fundamental mode at sky angle $\theta\simeq 1^{\circ}$ followed
by a succession of more or less harmonic secondary coherent
peaks at lower angles (which imply a nearly topologically flat
universe and a time-coherent creation of the sound waves), the
so-called causally generated oscillations, e.g. by cosmic string
network, were the result of fluctuations with sizes well inside
the horizon which never crossed it, and lead to a power spectrum
whose fundamental mode is shifted to angles smaller than
$\theta\simeq 1^{\circ}$ and whose secondary peak structure is
destroyed by temporal incoherence induced by causal random
forcing of the oscillators and ausence of common horizon
crossing.

Reliable observations of the power spectrum of CMB anisotropies
have only been made quite recently. Boomerang [20] and Maxima
[21] results seem now able to allow some discrimination between
the above contending models. Thus, the coincidence of a first
peak at $\theta\simeq 1^{\circ}$ in both experiments point
clearly in favour of inflation (see however Refs. [53,54]).
However, although the results seem clearly in favour of
whichever models where the perturbations crossed the horizon to
exceed it thereafter until last scattering, it is by no means so
clear that they may declare inflation as the only particular
mechanism which is responsible for the observed power spectrum.
In fact, the two experiments also agree in that the amplitude of
at least the second peak at $\theta\simeq 0.35^{\circ}$ is
dramatically smaller than that is predicted by inflation [55].

Inflation appears then to be in trouble when trying to explain
not only the amplitude of density fluctuations estimated by COBE
[19], as pointed out in Subsec. V.B, but also the spectrum of
CMB anisotropies measured by Boomerang [20] and Maxima [21].
Thus, the time could be up for the emergence of new alternatives
where perturbations were also forced to cross the horizon. In
what follows we show that Misner-brane cosmology might be one of
such alternatives. While the perturbations in our approach keep
the "good" property of horizon crossing, they possess two novel
properties which are worth remarking and might be of much
interest to justify experiment.

Firstly, since metric perturbations in this case can generally
be written as
\begin{equation}
h_{ab}=(n'+\alpha)\eta^2=(n'+\alpha)(n+\alpha)^4 ,
\end{equation}
with $n'\leq n$, fluctuations will cross the horizon not just at
a common time $\eta_c\simeq 1$, for $n'=n=1$, but also during a
following time interval $\bigtriangleup\eta$ which is
characterized by integral numbers such that $n'<n$. On the brane
at $\omega=0$, which has positive Newton constant $G_N >0$, the
perturbations smaller than the horizon size will then be
causally washed out eventually, so that one should expect the
interval for the time-incoherent horizon crossing
$\bigtriangleup\eta$ to be small, though nonzero. The resulting
situation would match what is predicted by inflation, except for
the involved time-incohence induced by a nonzero
horizon-crossing time interval which would partially destroy the
coherent structure of the secondary peaks. Secondly, on the
brane with $G_N <0$ at $\omega=\omega_c$, only those
fluctuations which are inside the horizon ($n' <n$) are able to
keep the oscillations stable. This is caused by inversion of the
gravitational forces and hence radiation pressure in the wells
[56]. Fluctuations exceeding the horizon size will then be
inexorably washed out. However, even subhorizon fluctuations on
the brane at $\omega=\omega_c$ will eventually cross the horizon
and then inexorably dissipate out, except possibly for the
residual fraction with the smallest values of $n'$, whenever the
automorphic parameter $\alpha$ and the minimum time $T_0$ take
on values that satisfy $\alpha\ln^{1/3}T_0^2\leq\eta_* ^{-1}$,
with $\eta_*$ being the recombination time. This constraint on
$\alpha\ln^{1/3}T_0^2$ is compatible with the amplitude of the
density contrast considered in Subsec V.B. Fluctuations
surviving inside the horizon at recombination on the brane with
negative $G_N$ for small $n'$ will also produce oscillations
with fundamental mode at $\theta\simeq 1^{\circ}$, and secondary
peaks whose structure is destroyed by temporal incoherence
induced by causal random forcing of the oscillators.

In summary, Misner-brane cosmology predicts quantum fluctuations
satisfying evolution equations in the radiation-dominated regime
of the form [50,57]:
\begin{equation}
[\Theta+\Psi](t_*)=[\Theta+\Psi](0)\cos(ks)
\end{equation}
\begin{equation}
v_{\gamma}=\sqrt{3}[\Theta+\Psi](0)\sin(ks) ,
\end{equation}
where $\Theta=\delta T/T$, $\Psi$ is the Newtonian potential,
$t_*=\int_0^{\eta_*}d\eta/a(\eta)$ is the conformal time at last
scattering, and $s=\int_0^{t_*}c_s dt$ is the sound horizon at
last scattering, with $c_s$ the sound speed and
$\Theta=\mp\Psi/2$, with the choice of sign depending on whether
perturbations on the brane at $\omega=0$ (upper sign) or on the
brane at $\omega=\omega_c$ (lower sign) are considered. Since
Misner symmetry identifies the two branes, relative to an
hypothetical observer who is able to travel through the fifth
direction, the perturbations on both branes should
simultaneously contribute the power spectrum for CMB
anisotropies [50,57]. Thus, relative to that observer and
$\omega=0$, this spectrum becomes:
\begin{equation}
C_{\ell}\simeq\frac{2}{\pi}\sum_{\omega=0}^{\omega_c}
\int\frac{dk}{k}k^3\left[\left(\Theta_{\omega}
+\Psi_{\omega}\right)\jmath_{\ell}(kd)
+v_{\gamma\omega}\jmath '_{\ell}(kd)\right]^2 ,
\end{equation}
in which the $\jmath_{\ell}$'s are spherical Bessel functions of
the first kind [33], and $d=t_0 -t_*$, with $t_0$ the current
conformal time.

For current observers who only are able to observe what is
happening on just the brane at $\omega=0$, the power spectrum
will reduce to
\begin{equation}
C_{\ell}\simeq\frac{2}{\pi}
\int\frac{dk}{k}k^3\left[\left(\Theta_{\omega=0}
+\Psi_{\omega=0}\right)\jmath_{\ell}(kd)
+v_{\gamma\omega=0}\jmath '_{\ell}(kd)\right]^2 .
\end{equation}
Obviously, the spectrum given by Eq. (5.26) produces the same
fundamental ($\ell\simeq 180/\theta\simeq 200$) and overtone
($\ell\simeq 500$, etc) acoustic modes as in the spectrum
predicted by simplest inflationary models [], but with the
rather remarkable difference of having quite smaller heights for
the secondary peaks. Although this prediction seems to be quite
compatible with Boomerang and Maxima results, it appears that
any conclusion will only be attainable from the future results
provided by MAP [58] and Planck surveyor [59] at lower angles.

\section{\bf Conclusions}
\setcounter{equation}{0}

Within the spirit of the Randall-Sundrum approach, we have
considered a five-dimensional apacetime whose metric satisfies
Misner symmetry, discussing the cosmological and gravitational
implications arising from the resulting brane-world model. After
reviewing the Misner-brane universe and analysing some of its
physical characteristics, a Misner-brane black hole scenario has
been constructed in which the conventional metric of a
Schwarschild or anti-de Sitter black hole is obtained when
neutral matter in the branes is allowed to collapse without
rotating. By using then a semiclassical approximation to the
Euclidean path-integral approach, we have calculated the quantum
state of the Misner-brane universe and the quantum effects
induced on its metric by brane propagation along the extra
coordinate. It has been seen that Misner symmetry requires that
the absolute value of the tension in the two branes be always
the same, and hence it follows that our brane scenario is fully
incompatible with the existence of any inflationary period in
the radiation dominated era. However, since communications
between distant regions which are outside the horizon can still
be done both causally and by means of CTC's through the fifth
dimension, the horizon and flatness problems can be solved in
our model.

Perhaps the most remarkable results in this paper be the
predictions of a scale-independent spectrum of density
fluctuations whose amplitude can, contrary to inflation, be
easily accommodated to the existing observational bounds.
Density fluctuations come here about as a result from the
existence of metric perturbations on the branes induced by the
above-alluded brane propagation along the extra direction, at
the earliest cosmological times. At later times, these
perturbations grow beyond the horizon during a nonzero time
interval on the two branes and give rise therefore to a power
spectrum of CMB anisotropies whose acoustic peaks are at exactly
the same small sky angles as in the spectrum corresponding to
the simplest inflationary models, but with secondary peaks whose
intensity is expected to be greatly diminished with respect to
the inflation-generated spectrum. This prediction seems to
conform to recent measurements by Boomerang and Maxima better
than those from inflationary models do.

\acknowledgements
The author thanks C.L. Sig\"uenza for useful discussions. This
work was supported by DGICYT under Research Project No.
PB97-1218.

\end{document}